\documentclass[12pt,reqno]{article}

\usepackage{latexsym}
\usepackage{amsmath}
\usepackage{amssymb}
\usepackage{amsthm}
\usepackage{amsxtra}
\usepackage{amsfonts}
\usepackage{mathrsfs}
\usepackage{fancyhdr}
\usepackage{amsbsy}
\usepackage{color}

\usepackage[utf8]{inputenc}
\usepackage[english]{babel}
\usepackage[T1]{fontenc}
\usepackage{makeidx}
\usepackage{graphicx}

\newcommand{\ba}{\begin{eqnarray}} 
\newcommand{\ea}{\end{eqnarray}}
\newcommand{\be}{\begin{equation}} 
\newcommand{\ee}{\end{equation}}

\numberwithin{equation}{section}

\makeindex

\title{\textbf{An algorithm-based introduction to the evolution of physical systems}}
\author{Emilio Balzano\footnote{Dipartimento di Fisica ``Ettore Pancini'', Universit\`a degli Studi di Napoli Federico II.
Istituto Nazionale di Fisica Nucleare. Email: balzano@na.infn.it},
Eliana D'Ambrosio\footnote{Dipartimento di Fisica ``Ettore Pancini'', Universit\`a degli Studi di Napoli Federico II. \,\,\,\,\,\,\,\,\,\,\,\,\,\,\,\,\,\,\,\,\,\,\,\,\,\,\,\,\,\,\,\,\,\,\,\,\,\,\,\,\,\,\,\,\,\,\,\,
Email: dambrosio@fisica.unina.it},
Rodofo Figari\footnote{Dipartimento di Fisica ``Ettore Pancini'', Universit\`a degli Studi di Napoli  Federico II. 
Istituto Nazionale di Fisica Nucleare.
Email: figari@na.infn.it} 
}
\begin{document}
\bibliographystyle{./references/ref}
\bibliography{test}
\makeatletter
\let\@orig@endthebibliography\endthebibliography
\renewcommand\endthebibliography{%
  \xdef\@kept@last@number{\the\c@enumiv}%
  \@orig@endthebibliography}
\newenvironment{thesitography}[1]
   {\def\refname{Sitography}
   \thebibliography{#1}%
    \setcounter{enumiv}{\@kept@last@number}%
}
  {\@orig@endthebibliography}
\makeatother
    
\maketitle

\begin{abstract}
\noindent
We outline  an unified introduction to the evolution equations of classical and quantum systems intended for a high school students audience.
The attempt consists in circumventing the lack of mathematical knowledge with the use of simplified forms of  numerical analysis. The aim is to allow students to approach  theoretical features as well as computational aspects of the evolution equations.
In particular, the possibility to compute and analyze approximate solutions of the dynamical laws of classical, stochastic and quantum mechanics enables to highlight the distinctive role played by probability in different contexts. 
As computer support for numerical analysis we selected the spreadsheet. It  is a work environment usually presented in high school and it is an ideal tool for an intuitive approach to numerical computation through recursive algorithms.
The proposal was presented to  an audience composed by students of the course of Didactics of Physics and of  high school Science teachers.

\vspace{.3cm}
 \end{abstract}

\section{Introduction}
In the last years the Italian Ministry of Education have been recommending that quantum mechanics and relativity should be part of any high school physics curricula \cite{Indire}. As it is very well known, however, there are many peculiar obstacles to overcome in order to comply with those recommendations. In particular for quantum mechanics, the project faces obvious difficulties: on one hand, the impossibility of experiencing the quantum world in everyday life makes it difficult to use guided presentations of its phenomenology. \\
On the other hand, a not merely informative introduction to quantum mechanics requires  advanced mathematical tools (complex numbers, ordinary and partial differential equations, etc.), generally unknown to high school students.

\noindent
In the volume ``Innovations in Science and Technology Education'' edited by UNESCO,  Svein Sj\o berg analyzes the status of science and technology education and the way science and the scientists' work are perceived by young people \cite{Sj}.  \\
In a section devoted to the possible reasons of the ``disenchantment with science and technology''  the author criticizes many actual curricula and textbooks  which, in his opinion,  ``are overloaded with facts and information at the expense of concentration on a few ``big ideas and key principles'' often leading to ``...rote learning without any deeper understanding.''\\
Effectively,  physics and in particular quantum physics often appear to students as an arid enumeration of laws or a list of fragmentary pieces of information related to each other only by the chronological order of their discovery. In fact, the extraordinary power of a physical theory in giving a detailed description of some aspect of the real world is made evident  by the analysis of qualitative and quantitative features of the solutions of  the evolution equations. If the skills for performing such an analysis are lacking, students are deprived of the astonishment resulting from the verification of the peculiar effectiveness of the theory.\\
\noindent
In many contributions to the previously cited book the focus is also on the role of technological applications as a resource to outline the scientific approach to the comprehension of physical world as characterized "not just about knowing but also about doing and making things work."
Both new technological tools and conventional software (i.e. database, spreadsheet, graphical programs etc) are presented as key elements to develop teaching and learning together with modeling, visualization and simulation of processes.\\
\noindent
The latest proposals in response to students difficulties are in fact heading in this direction. Regarding quantum physics, new teaching strategies suggest introductions to basic concepts given with the help of dedicated softwares and of multimedia presentations \cite{Muller-Wiesner} \cite{Zollman} \cite{Kohnle}, \cite{VQM}, \cite{Thaller} with the attempt to bridge the gap between the abstract formalism of quantum mechanics and the qualitative understanding necessary to explain and predict physical phenomena. \\
\noindent
Other approaches,  always in the direction of  ``about doing and making things work",  demonstrates how computational physics, beyond being an effective way to find approximate solutions to specific problems, can help students to understand basic concepts in physics \cite{Vistnes}. As an example we want to mention the proposal of Hugdal and Berg \cite{Hugdal} that describes the quantization procedure in an intuitive way and presents a numerical algorithm to find solutions to the time-independent Schr\"odinger equation for several one dimensional potentials.\\
\noindent
Our idea is to simplify the computational methods to adapt them to computer skills possessed by high school students. In this paper we present an approach to numerical computation through recursive algorithms that will be implemented in a spreadsheet.
We chose it as computer support because it is largely used in school contexts and because we consider it an ideal environment for an intuitive approach to numerical solutions of the differential equations that govern a theory. 
In this way,  without having previous knowledge of programming languages, students can easily implement the algorithm and focus their attention on dynamical features of system's evolution. 
More precisely, we consider a discretized space-time where the evolution equations become recurrence relations that students are first required to handle with the use of a spreadsheet. Given any initial condition,  solutions at any future time are computed and graphically represented.  The continuous space-time limit is then qualitatively obtained looking at the solutions on a scale where discretization become unnoticeable, avoiding all mathematical details. 
In this way we suggest an unified methodology to analyze the evolution laws of classical, stochastic and (simple) quantum systems. 

\noindent
In our opinion, there are some advantages in such a presentation which are worth mentioning:

\begin{itemize}
\item
the dynamical laws are not simply stated but analyzed on the basis of their effectiveness in modeling the evolution of real world systems;
\item   
students are enabled to autonomously examine dynamical features  of complex systems whose study is generally considered too complicate to be part of a physics course in high school as well as in the college level general physics courses ( e.g.  many body gravitational systems,  large amplitude pendulum oscillations, stochastic dynamical laws, quantum wave packet evolution etc.); 
\item 
students are provided with some preliminary skill for examining how solutions depend on the initial conditions and on dynamical parameters; 
\item 
the possibility to compute and analyze approximate solutions of the evolution laws allows to make a first comparison between classical, stochastic and quantum world, to highlight the distinctive role played by probability in different contexts and to appreciate to what extent classical categories lose their meaning in the quantum context.

\end{itemize}
\noindent
In the following sections we detail the discrete version of the evolution equations we consider. 
We want to point out that our aim is not to counterpose a discrete mathematics to a continuos one. The theoretical necessity to give a rigorous meaning to the continuos limit of the evolution equations should be discussed and clarified. In particular, it should be suggested that this step requires new mathematics. \\ 
On the other side, it is clear that the the physical content and the explanatory power of a theory can be acquired from a discretized space-time formulation of the equations whenever approximate solutions are explicitly computable. In this context, we want to mention the book \cite{Russo}  where the authors apply a similar approach to analyze qualitative and quantitative features of the evolution  of classic dynamical systems  with particular emphasis to the onset of chaotic behaviors.\\
In our project, in particular,  we will use the simplest form of discretization of differential equations, without entering any stability problem of the numerical computation.  We will come back to this problem in the concluding section.

\noindent
As a final remark,  we share the view that is supported by an Italian research  \cite{Levrini} which, to contrast the conceptual fragmentation (further cause of difficulty in understanding quantum physics) highlights, in the introductory phase of quantum physics teaching, the importance of a presentation of phenomenology that relates the experimental aspects with the formalism of a theory without excluding epistemological aspects, as well as  it emphasizes the productivity of ``The ``game'' of modelling'' and of comparing the quantum models with ``the models already encountered by the students during the study of other physics topics''.

\section{Classical Dynamic}

\noindent
Newton's idea consisted in attributing preliminarily a cause to the motion of bodies : the ``forces'' acting on them. Then to detail quantitatively their effects. According to the laws of motion, the forces,  expressing the interaction between bodies, generate velocity changes. He realized that a large class of motions (the "natural ones") could be explained with the single hypothesis of the existence of the gravitational force acting among material bodies \cite{Newton}.\\ 
In this way, he was able to make available an unified explanation of the keplerian orbits of planets, of the free fall motion of bodies close to the earth surface, of the tides, of the small oscillations of the pendulum etc. 
\\
In order to find solutions to the dynamical equations of motion Newton was faced with the problem to reconstruct, starting from its initial position and velocity, the successive  positions and velocities of a material body as a result of a continuous infinity of velocity variations of the body itself. 
In fact, according to Newton:  ``Absolute, true and mathematical time, of itself, and from its own nature flows equably without regard to anything external''. The flow of time is then continuous, uniform, eternal, unrelated to outer space, and exists regardless of its measurement.

\noindent
The two step procedure conceived by Newton to compute solutions of the dynamical equations, used in particular in the analysis of the "two-body problem",  took later the name of ''Newton's Diagram'' \footnote{\footnotesize {The term, used with different meanings in different contexts, was utilized in the sense detailed below by Feynman in his lecture on the planetary motion later published in \cite{Pianeti}. }} and can be summarized as follows:
\begin{itemize}
\item the variation of the material body velocity during a ``short interval of time $\Delta t$ '' is proportional to the impulse of the force applied to the body during the same interval.  $\Delta t$ should be chosen short enough to guarantee that the force is not changing much during the time interval in such a way that the impulse can be well approximated by the force at the beginning of the interval times $\Delta t$.
\item The trajectory is obtained as a geometric limit of the broken line obtained considering the motion linear and uniform  during each interval.
\end{itemize}
Essentially, this protocol consists in solving for ``short $\Delta t$ ''  the following pair of equations:
\begin{equation} \label{New}
\left \{\begin{array}{ll}
\mathbf{x}(t+\Delta t)-\mathbf{x}(t)=\mathbf{v}(t)\cdot \Delta t \\
\mathbf{v}(t+\Delta t)-\mathbf{v}(t)=\mathbf{F}(\mathbf{x}(t),t)/m \cdot \Delta t
\end{array} \right.
\end{equation}
where the a-priori known $\mathbf{F}(\mathbf{x}(t), t)$ is the force acting on the pointlike body at instant $t$ (when it is in the position $\mathbf{x}(t)$)
\footnote{\footnotesize{Notice that a possible dependence of the force on the velocity  (for example a friction force)  is not going to complicate the computational procedure used to find solutions of \ref{New}.}}.

\noindent
The following vector identities expressing the total displacement and the total velocity variation of the point particle as a sum of position and velocity changes in each time interval $[t_{i-1} , t_{i}]$ (no matter how coarse or fine  the subdivision of the total time interval is)   are the only required kinematic prerequisites
  
\begin{equation}\label{kin1}
 \vec{r}( t_{N}) - \vec{r}( t_{0}) = \sum_{i=1}^{i=N} \left( \vec{r}( t_{i}) - \vec{r}( t_{i-1})\right) = \sum_{i=1}^{i=N} \frac{ \vec{r}( t_{i}) - \vec{r}( t_{i-1})}{( t_{i}  -  t_{i-1})} ( t_{i}  -  t_{i-1})
= \sum_{i=1}^{i=N} \vec{v}_{t_{i-1},t_{i}}  ( t_{i}  -  t_{i-1})
\end{equation}
the ``average velocity" between times $ t $ and $ t '> t $ being defined as
 $\vec{v}_{t,t'} \equiv\frac{\vec{r}( t')  - \vec{r}( t)}{ t'  -  t}.$\\
In the same way, the total velocity variation is expressed as follows
\begin{equation}\label{kin2}
   \vec{v}( t_{N})  -  \vec{v}( t_{0}) =  \sum_{i=1}^{i=N} \left( \vec{v}( t_{i}) - \vec{v}( t_{i-1})\right) = \sum_{i=1}^{i=N} \frac{ \vec{v}( t_{i}) - \vec{r}( t_{i-1})}{( t_{i}  -  t_{i-1})} ( t_{i}  -  t_{i-1}) = \sum_{i=1}^{i=N} \vec{a}_{t_{i-1},t_{i}}  ( t_{i}  -  t_{i-1}) \,\,\,\,\,
\end{equation}
where the ``average acceleration'' between times $ t $ and $ t '> t $ is defined as
\[\vec{a}_{t,t'} = \frac{\vec{v}( t')  - \vec{v}( t)}{t'  -  t}.\]

\noindent
Within a discretized time scheme one can easily define work, kinetic and potential energy as well as conservativeness of harmonic and gravitational forces.   The dynamical law (\ref{New}) and the kinematical identities (\ref{kin1}),(\ref{kin2}) then allow to prove conservation of the total energy apart from terms that become negligible when the time step of temporal discretization is small.

\noindent
We want to emphasize  concreteness, easiness and effectiveness of the discrete approach coupled with an affordable way to compute numerical solutions to equation (\ref{New}).  However, it is clear that, as long as there is no absolute minimal time interval value (as it is the case when a continuous model of time is assumed)  a rigorous theory is missing until the value of $\Delta t $ remains unspecified.

\noindent
Implementing the recurrence (\ref{New}) into a spreadsheet, it is possible to compute numerically the point mass  motion as a function of time. More precisely, when position and velocity (and, in turn, the force) are known at the initial time, the recurrence equations return  position and  velocity at any time $ t_{i}$.
Each computational step consists in fact in the copy and paste  of the  previous line, where the recursion formulas are written via relative references (apart from fixed parameters appearing as absolute references).\\
In the following, we show how the procedure outlined  above applies to few examples of classical systems which are  considered, in general, too complicate to be presented in an elementary physics course. The aim is to show that the acquisition of qualitative and quantitative understanding of relevant features of the evolution of complex system is surely within high school students' reach.

\vspace{.2cm}
\noindent
Newton used the law of universal gravitation together with the second law of motion to analyze the two body gravitational problem.  He characterized all possible orbits and verified the validity of Kepler's laws of planetary motion. In the `` Principia ", he made also the first attempt to deal with the gravitational three body problem in order to calculate the effects of the Sun on the Moon's motion around the Earth.  Long after, Euler  analyzed  a simplified version of the dynamical model successively denoted by Poincar\'e as the "circular restricted three body problem". In this model three pointlike bodies interact via gravitational forces; one of the body has a mass which is negligible with respect to the masses of the other two and the two ``heavy'' bodies follow circular orbits around their center of mass. The circular restricted three body problem marked the birth of perturbation theory in celestial mechanics which allowed to understand and compute the secular variations of planet orbits and started the investigations about the stability of the solar system. The achievements of celestial mechanics were undoubtedly the greatest and most astonishing success of newtonian mechanics. \\
In the way outlined above, it is possible to  investigate numerically the evolution of a simplified circular restricted three body planetary system. For the sake of simplicity, we assume 
\begin{itemize}
\item that the heaviest body (the sun) has a mass $M_{S}$ so large to be subjected to negligible acceleration. Its fixed position will be then taken as the origin of the cartesian coordinate system inside which the motion of the two planets is described.  
\item A planet of large mass $M_{e}$ is assumed to follow a circular orbit around the sun. Its position at time $t$ is denoted by $\mathbf{x}_{e}(t) = (x_{e}(t), y_{e}(t))$
\item A second  planet, of mass $m$ negligible with respect to the first, has an initial velocity in the plane containing the sun and the two planets at the initial time in such a way that its motion will always develop on that plane.  The initial distance of the light planet will be taken smaller than the radius of the orbit of the heavy one and for this reason the second planet will be referred to as the inner planet (and the first as the outer planet). Its coordinates will be denoted by $\mathbf{x}_{e}(t) = (x_{e}(t), y_{e}(t))$
\end{itemize}
\noindent
Under these assumptions it is possible to use (\ref{New}) to examine  the motion of the light planet subject to the gravitational action of the Sun and of the outermost planet for various initial conditions and mass ratios. 

\begin{figure}  [ht]
\includegraphics[width=170mm]{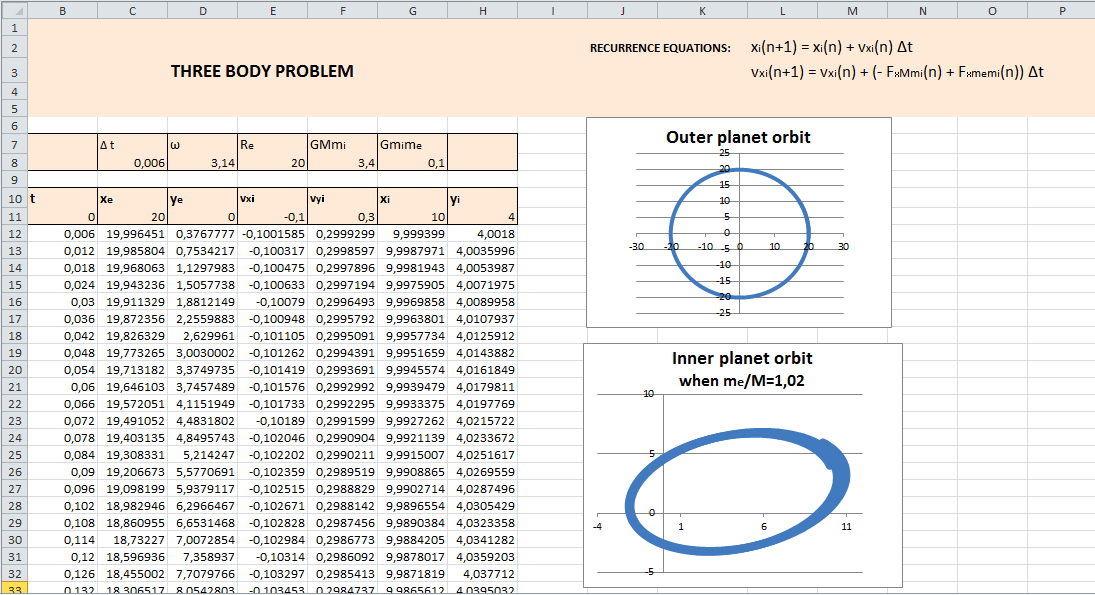}
            \caption{\emph{An example  of a worksheet showing  computation and plots of the planet orbits in a simplified circular restricted three body planetary system.} }
            \label{tre}
    \end{figure}

\noindent
In the following example the components of the outer planet positions are  taken to be
\[x_e(t)= R_e cos\,\omega_{e} t;\,\,\,\,\,\,\,\,\,\,\, y_e(t)=R_e sin\, \omega_{e} t \] \\
The (\ref{New}) reeds in this case
\begin{equation} 
\left \{\begin{array}{ll}
\mathbf{x}(t+\Delta t)-\mathbf{x}(t)=\mathbf{v}(t)\cdot \Delta t \\
\mathbf{v}(t+\Delta t)-\mathbf{v}(t)=(\mathbf{F}_{M_{S}m}(\mathbf{x}(t))+\mathbf{F}_{M_{e}m}(\mathbf{x}(t),t))/m \cdot \Delta t
\end{array} \right.
\end{equation}

\noindent
where the $x$ components of the forces on the inner planet due respectively to the Sun and to the outer planet are
\begin{equation}
\begin{array}{ll}
F_{M_{S}m}(\mathbf{x}(t))=\frac{GM_Sm}{[(x^2(t)+y^2(t))]^{\frac{3}{2}}}\cdot x(t)\\
F_{M_em}(\mathbf{x}(t),t)=\frac{GM_em}{[(x_e(t)-x(t))^2+(y_e(t)-y(t))^{2}]^{\frac{3}{2}}}\cdot (x_e(t)-x(t)).\end{array}
\end{equation}
\renewcommand{\figurename}{Fig.}
The $y$ component can be easily obtained replacing $x$ with $y$. \\
Implementing these equations in the spreadsheet (\figurename\ref{tre}), it is possible to examine the motion of the inner planet, for various values of the dynamical parameters (the ratios $M_e/M_S$ and $m/M_S$) and for different initial conditions. In particular, students should be able to analyze the onset of a chaotic behavior of the system and the strong dependence on the initial conditions in the chaotic regime. Notice that an implicit space discretization is implied by the choice of the maximum number of decimal places allowed in the spreadsheet.

\vspace{.3cm}
As a second example of mechanical system that can be thoroughly examined via numerical computation  implemented in a spreadsheet we want to mention the oscillatory motion of a pendulum, for generic initial conditions, in presence of viscous friction.
Indeed, the numerical computation procedure to search solutions to (\ref{New})  does not present any relevant difficulty even  in the case of non linear forces. It is in fact possible to analyze  cases of oscillations driven by forces whose elastic coefficients depend non linearly by the displacement. \\
The recurrence equations for a pendulum in presence of viscous friction,  are
\begin{equation}
\left \{\begin{array}{ll}
\theta(t+\Delta t)=\theta (t) +\omega(t) \Delta t\\
\omega(t+\Delta t)=\omega (t) -\alpha(t)-\beta\omega (t)\Delta t\end{array} \right.
\end{equation}
where $\alpha(t)=(g/l)\cdot\sin\theta(t)$ is the angular acceleration,  $\omega(t)$ is the angular velocity and $\beta$ is the damping coefficient. \\
Students can examine several features of the evolution. In particular it is possible to investigate the dependence of the period $T$ on the initial conditions and on the dynamic parameters and find out that
\begin{itemize}
\item{} the period shows negligible changes for small oscillation amplitudes ($\theta<5^o$).
\item{} Isochronism is lost for large oscillation amplitudes.  
\item{} The period depends in a negligible way on dumping: even if the oscillation amplitude decreases, the period remain essentially constant.
\end{itemize}

\noindent
With the same procedure outlined above it is possible to investigate the dynamics of classical fields. With respect to the previous cases, the numerical analysis of the evolution equations of a spatially continuous system requires also an explicit discretization of the spatial coordinates. Classical fields become functions on a  discrete space-time lattice.\\  
The longitudinal vibrations of an elastic string  represent the simplest example of a deformable body dynamics. A physical model of a discretized one dimensional elastic string can be realized as a chain of $N$ point masses which interact with their neighbours through  massless springs of elastic constant $k$ and rest  length equal to the spatial lattice spacing $ \Delta x$.  Denoting with $s(j \Delta x, n \Delta t)$ the displacement with respect to the equilibrium position  of the $j^{th}$ mass at time $n \Delta t$, the force acting on that mass $m$ is 
\begin{equation}
{\begin{array}{ll}
{F_{j \Delta x, n \Delta t}= k[s((j+1)\Delta x, n \Delta t)-s(j \Delta x, n \Delta t)]-k[s(j \Delta x, n \Delta t)-s((j-1)\Delta x, n \Delta t)] }\\
\\
{= k[s((j+1)\Delta x, n \Delta t) -2 s(j \Delta x, n \Delta t) +  s((j-1)\Delta x, n \Delta t)] }\end{array}}
\end{equation}
The equations of motion become

\begin{equation} 
\left \{\begin{array}{ll}
{s( j \Delta x, (n+1) \Delta t)-s( j \Delta x, n \Delta t)=v(n \Delta t)\cdot \Delta t} \\
v((n+1) \Delta t)  -v (n \Delta t) = \frac{1}{m} \,\,F_{j \Delta x, n \Delta t} \cdot \Delta t
\end{array} \right.
\end{equation}
\noindent

\noindent
with $j\in -N/2,...,N/2$ e $n\in \mathbb{Z}$. Given the displacements $s( j \Delta x)$  and the velocities $v(n \Delta x)$  at time $t=0$ the solution of the recurrence equations return displacements  and velocities of any point mass at any time $n \Delta t$. \\
Students are required to examine the way the evolution depends on the initial conditions and on the dynamical parameters and  to investigate traveling and standing wave solutions of the wave equation. 

\section{Stochastic Dynamic}
From the beginning of the last century, starting from the Einstein \cite{Einstein1},  Smoluchowski  \cite{Smo} and Perrin \cite{Perrin} analyses of brownian motion
models of stochastic evolution had great relevance in theoretical physics. The history of the scientific debate about the causes of brownian motion and of the role it played in the success of the molecular kinetic theory of matter is a fascinating subject worth learning in high school (see for example \cite{Gora} \cite{Mayo})

\noindent
Probability entered the theory to account for some practical limitations. Among them  
\begin{itemize}
\item
the impossibility to characterize at any level of accuracy the mechanical state of the brownian particle environment, which was assumed to be made up of a huge number of particles,
\item 
the unworkable target to specify the exact overall effect on the brownian particle of an almost continuous sequence of collisions. 
\end{itemize}
\noindent
Notice that are nowadays available high quality physics simulation softwares, based on object oriented programming, that students can easily use to simulate and visualize the motion of a massive large particle triggered by the elastic collisions with a large number of light small particles (see for example the animation in the wikipedia voice of the brownian motion \cite{wiki}). \\
In order to go from phenomenology and experimental results to the evolution equation for the position probability density it is obviously  required a previous knowledge of some basic elements of probability theory such as joint and conditional probability and Bayes' theorem \cite{Bao}. 

\vspace{.3cm}
\noindent
In the republication in English  "Investigations on the Theory of the brownian movement" \cite{Einstein} Einstein's assumptions appear in the following way:
\begin{itemize}
\item ``a force acts on the single particles, which force depends on the position but not on the time. It will be assumed for the sake of simplicity that the force is exerted everywhere in the direction of x axis''
\item ``each single particle executes a movement which is independent of the movement of all other particles''
\item ``the movements of one and the same particle after different intervals of time must be considered as mutually independent processes''.
\end{itemize}
He considers the number of particles per unit volume and infers the distribution of particles at time $t+ \Delta t$ from the distribution at time $t$ according to the assumptions reported above. \\
\noindent
In a one-dimensional lattice Einstein assumptions amount to state that the particle is undergoing a simple symmetric random walk: at each time step the particle have to jump to one of the two neighboring site with equal probability $1/2$. As a consequence, the position probability at  time $n+1$ depends only on the position probability at the previous instant $n$.\\
 In detail, 
the probability that the particle is in the lattice point $i$, at the time step $n+1$, is the sum of the probabilities of the two independent events: either the particle was at the lattice site $i-1$ at time $n$ and jumped to the right or the particle was at the lattice site $i+1$ at time $n$ and jumped to the left. Given the assumption of equal transition probabilities, the evolution equation is 
\begin {equation}\label{Brow}
 P_i(n+1)=\frac{1}{2}\, P_{i-1}(n)+\frac{1}{2}\,P_{i+1}(n).
\end{equation}
In a form which is more similar to the diffusion equation (\ref{Brow}) reads

\begin{equation} \label{diff}
P_i(n+1) - P_{i}(n) = \frac{1}{2}\, \left[ P_{i-1}(n) + \,P_{i+1}(n) - 2\,P_{i}(n) \right]
\end{equation}

\begin{figure} \centering
\includegraphics[width=100mm]{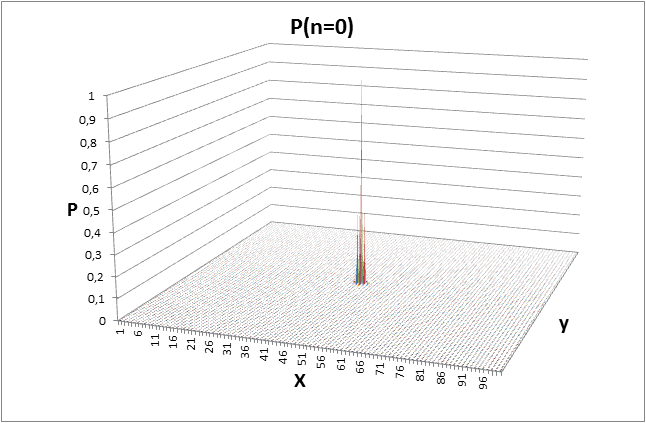}
\includegraphics[width=80mm]{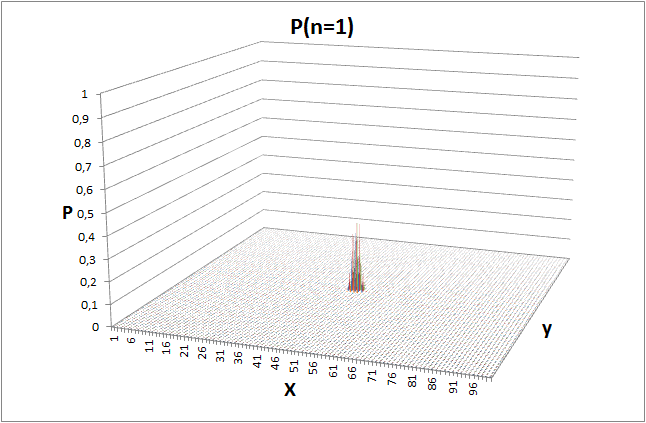}
\includegraphics[width=80mm]{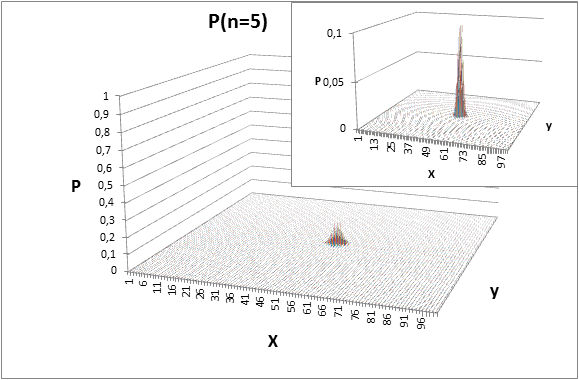}
\includegraphics[width=80mm]{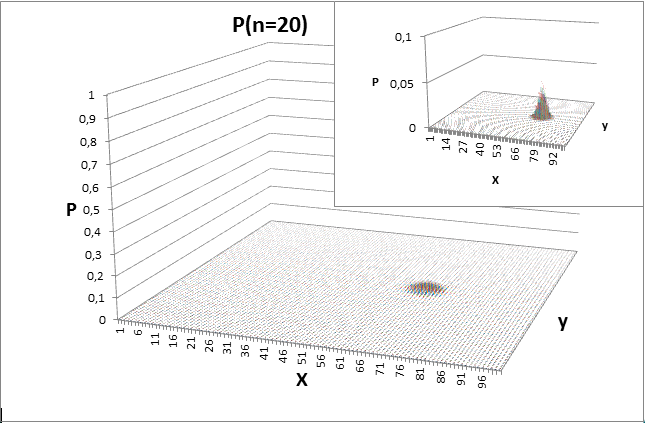}
\includegraphics[width=80mm]{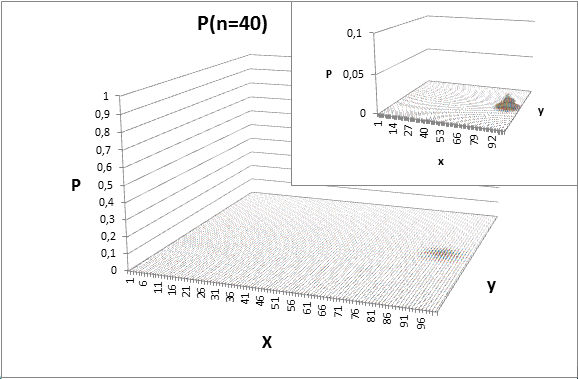}
        \caption{\emph{Evolution of the position probability of a particle performing a brownian motion superimposed to a classical drift.} }
        \label{drift} 
 \end{figure}

\noindent
The recurrence equation (\ref{diff}) is easily implemented in the spreadsheet. Given an initial position probability distribution, its solution  returns the brownian particle position probability at any time $n$. In diffusion of a gaussian initial distribution we plotted the solution of (\ref{diff}) for a gaussian initial distribution in a 3-d plot (probability density versus time and position). It is possible to observe the typical spread of the gaussian variance.\\
It also possible to investigate stochastic evolutions where a brownian motion is superimposed to classical drifts given by simple ``vector fields'' ({\figurename\ref{drift}}).
 
\begin{figure}[ht] \centering
\includegraphics[width=150mm]{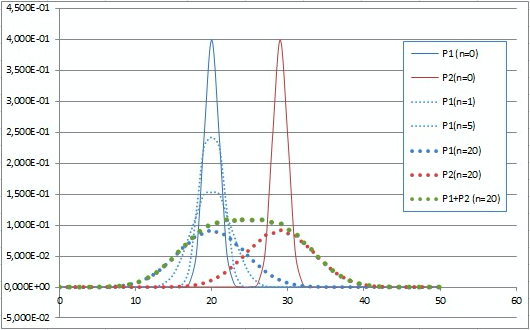}
        \caption{ \emph{The position probability of a brownian particle is given as the sum of the two independent evolutions.}}
         \label{due}
\end{figure}

\vspace{.3cm}
\noindent
Particularly important for the comparison with the evolution of the position probability of a quantum free particle is the case when the initial distribution consists of two distinct ``bumps'' respectively at the right and the left of the origin. In the brownian particle case the two bumps evolve independently and the probability to be anywhere at any successive time is given as the sum of the two independent evolutions ({\figurename\ref{due}}).

\section{Quantum Dynamic}
The number of books and papers in Science Education aimed at teaching and popularizing quantum physics at high school level has been growing rapidly in the last decades (see for example \cite{Arroyo} and \cite{Ghi}). For an exhaustive reference list and for a point of view we wholeheartedly  agree with we want to mention \cite{Levrini}, an article we already referred to in the introduction. In the paper, the authors start pointing out how hyper-simplification in  teaching modern physical theories run the risk to  corrupt  the process of learning Physics. Successively, they proceeded to describe a teaching proposal grounded on two suggestions:
\begin{itemize}
\item the importance to introduce students to the main aspects of the historical debate among the founding fathers of the theory about the foundation of quantum mechanics 
\item the need to present key experiments along with elements of the formalism that enable to comprehend the experimental outputs inside the theoretical model.
\end{itemize}
The latter aspect is addressed considering experiments ``\`a la Stern and Gerlach'' theoretically analyzed in the finite dimensional spin state space, in the way proposed by Pospiech \cite{Posp}. \\
In the following, we suggest an alternative way to approach the formalism presentation based on the Schr\"odinger's "Wave Mechanics". In line with what was done in the previous sections we formulate the Sch\"odinger equation in a discretized space-time as a (kind of) wave equation for a two dimensional vector field (the real and the imaginary part of the wave function). The history of the attempts, made during the early stage of quantum mechanics, to understand the role of the wave function $\Psi$ (Schr\"odinger, Born) should then guide to formulate the way to extract the expected statistics in position or momentum measurements. \\
Main motivations of the choice are:
\begin{itemize}
\item by now, a large number of good quality multimedia presentations of the two slit experiment for matter waves are available (see for example \cite{film}, \cite{applet1}, \cite{applet2}). The experiment shows in a clear way that equal experimental initial settings result in different impact positions and at the same time that the impact statistics in repeated experiments has ``something to do'' with a wave propagation phenomenon (for a clear and fascinating introduction to the two slit experiment, see \cite{Bell}).
In our opinion there is no better way to convey the ontological character of probability in quantum phenomena and to convince students of the necessity to find out the ``right'' wave equation carrying all the information about the statistics of any possible measurement.
\item The particle-wave duality is a hard to perceive concept whose origin goes back to the early days of quantum mechanics. In our opinion, it has always  been a source of uncertainty and disorientation. On the contrary,  a naive point of view maintaining that particle means simply that position and momentum are well defined observables and that a wave is the carrier of all available statistical information about the quantum particle observables seems to us the (interpretation-free) less confusing way to present to a high school students audience the way a quantum particle state evolves.
\end{itemize}

\noindent
As it was previously done, we will assume that space is a large lattice with a very small lattice spacing. 
In summary, the theoretical assumptions concerning the state space, the expected statistics of measurement results and on the ``right'' state time evolution assume the following form in a one dimensional lattice with $2N+1$ sites:
\begin{itemize}
\item Kinematics: the state of a particle at any time $n$ is specified by the association, at each point $i$ of the lattice, of a vector with two components: $\displaystyle\Psi (n) \equiv\left\{\vec{ \psi_j}\right\}_{j=-N}^{N}(n)\equiv\left\{ \psi_{j,x}, \psi_{j,y} \right\}_{j=-N}^{N}(n) \equiv \{x_j(n),y_j(n)\}_{j=-N}^{N}.$  \\
To the physical observables position and momentum are associated  the operations $Z$ and $P$ in the state space

\begin{equation}
Z \Psi (n) = \left\{\overrightarrow{ (Z\Psi)_j}\right\}_{j=-N}^{N}(n) = \{j\,x_j(n),j\,y_j(n)\}_{j=-N}^{N} 
\end{equation}

\begin{equation} \label{moment}
P \Psi (n) = \left\{\overrightarrow{ (P\Psi)_j}\right\}_{j=-N}^{N}(n) = \{\Delta_{j} y(n),\, - \Delta_{j} x(n)\}_{j=-N}^{N}
\end{equation}
with $\displaystyle \Delta_{j} x = \frac{x_{j+1} - x_{j}}{a}$ and $\displaystyle \Delta_{j} y = \frac{y_{j+1} - y_{j}}{a}$ and where $a$ denotes the lattice spacing.\\
When the state at time $n$ is $\Psi(n)$, the following real numbers are associated to position and momentum 
\begin{equation}
\left(\Psi (n), Z \, \Psi (n)\right) \equiv \sum_{j=-N}^{N} \left(\psi_{j,x}(n)\,(Z\psi)_{j,x}(n) + \psi_{j,y}(n)\,(Z\psi)_{j,y}(n)\right) = \sum_{j=-N}^{N} j\,(x^{2}_{j}(n) + y^{2}_{j}(n)) \nonumber
\end{equation}

\begin{eqnarray}
\left(\Psi (n), P \, \Psi (n)\right) &\equiv &\sum_{j=-N}^{N} (\psi_{j,x}(n)\,(P\psi)_{j,x}(n) + \psi_{j,y}(n)\,(P\psi)_{j,y}(n)) \nonumber\\
& = &\sum_{j=-N}^{N} j(x_{j}(n)\,\Delta_{j} y(n) - \,y_{j}(n) \Delta_{j} x(n))  \nonumber
\end{eqnarray}

\item Statistics of measurement outputs:   the probability to find the particle at the point $i$ of the lattice, at
time $n$ is 
\[ P_j(n)=|\vec{\psi}_j(n)|^2= {x_j}^2(n)+y_j^2(n).\]
Since the particle is necessarily somewhere, 
it must be true that $\sum_{j=-N}^{N}(x_j^2(n)+y_j^2(n))=1$.

In particular, the average expected value of the particle position is given by
\[ \langle Z \rangle = \left(\Psi (n), Z \, \Psi (n)\right)= \sum_{j=-N}^{N} j\, (x_j^2(n)+y_j^2(n)) \]
and the position mean squared error 
\[ \langle (Z - \langle Z \rangle)^{2} \rangle  = \langle Z^{2} \rangle  -\langle Z \rangle^{2}= \sum_{j=-N}^{N} j^{2}\, (x_j^2(n)+y_j^2(n))  - \langle Z \rangle^{2} \]
The average expected value of momentum $p = m\,v$ is 
\[ \langle P \rangle = \left(\Psi (n), P \, \Psi (n)\right) = \sum_{j=-N}^{N}  (x_j(n) \Delta_{j}\,y(n) - y_j(n) \Delta_{j}\,x(n)) \]
and the momentum mean squared error
\[ \langle P^{2}  - \langle P \rangle^{2}\rangle = - \sum_{j=-N}^{N}   \Delta^{2}_{j}\,y(n) + \Delta^{2}_{j}\,x(n)) - \langle P \rangle^{2}\]

\item
The state evolution: The components of the state vector evolve in time according to the Schr\"odinger equation:
\begin{equation}\label{Sch}
\left\{ \begin{array}{lll}
x_j(n+1)-x_j(n)=[-y_{j+1}(n)-y_{j-1}(n)+2y_j+ V_j y_j(n)]\Delta t\\
y_j(n+1)-y_j(n)=[+x_{j+1}(n)+x_{j-1}(n)-2x_j+ V_j x_j(n)]\Delta t\\
\end{array} \right.
\end{equation}
where $V_{j}$ is the value assumed in the point $j$ by the potential of the conservative force acting on the quantum particle.
\end{itemize}
 Let us list few immediate consequences of the assumptions  
 \begin{description}
 \item{a)} $\Phi = \{\alpha \delta_{ij} , \beta  \delta_{ij} \}_{i=-N}^{N},\,\,\,\, \alpha^{2} + \beta^{2} = 1$ is the state of a particle which is with certainty in the point $j$. In the state $\Phi$ momentum has a maximal mean square error (equal to 2). 
 \item{b)} $\Psi = \{\frac{\alpha}{\sqrt{2N+1}} , \frac{\beta}{\sqrt{2N+1}} \}_{i=-N}^{N},\,\,\,\, \alpha^{2} + \beta^{2} = 1$ is the state of a particle which has with certainty zero momentum. The corresponding position distribution is uniform implying that position has maximal uncertainty in the state $\Psi$. 
\item{c)} the particle state $\{x_j\,,\,y_j\}_{j=-N}^{N}$ obtained starting from $(x_N = \frac{1}{\sqrt{2} \sqrt{2N+1}}\,,\,y_N = \frac{1}{\sqrt{2} \sqrt{2N+1}})$ and rotating by a small angle $\theta$
\begin{eqnarray}
x_{j+1}& = & \,\,x_{j} \cos\theta + y_{j} \sin \theta\,\,  \simeq \,\, x_{j}  + y_{j} \theta \nonumber \\
y_{j+1}& =  &-x_{j} \sin \theta + y_{j} \cos \theta \,\simeq \, -x_{j}  \theta + y_{j} \nonumber
\end{eqnarray}

 is the state of a particle having with certainty momentum $\theta$ as it is clear from (\ref{moment}). Notice that the position probability distribution in this state is again uniform.
 \item{d)} There are no states for which the particle has, with certainty, a precise value of position and momentum.
 \item{e)} A state describing a particle with zero value of momentum does not evolve in time if no force is acting on it ($V_{i} = 0\,\,$ for all $i$).
 \end{description}

\subsection {Quantum dynamics in some simple cases}
A theoretical description of the two slit experiment based on the ``axiomatic''  formalism introduced above should convince students that, from one side, the postulates are ad hoc assumptions needed to explain the observed interference of ``matter waves''. On the other side, the statistics of possible outputs of measurements, obtained via the auxiliary vector $\Psi$, are at odds with any classical description of a stochastic evolution.    \\
The state vector evolves like a ``wave of possibilities'' but it is not at all a probability density. In the words of Bell: \emph{ ``What is it that waves in wave mechanics? In the case of water waves it is the surface of the water that waves. With sound waves the pressure of the air oscillates. Light also was held to be a wave motion in a classic physics.[...] In the case of the waves of wave mechanics we have no idea what is waving and do not ask the question.''}

\noindent
 The consequences of the linearity of the Schr\"odinger equation together with  the superposition principle can be made clear and explicitly shown in simple cases. 
 In particular, in a one dimensional lattice we consider a free quantum particle in an initial state $\Psi$ that is a normalized superposition of two states, $\Psi^{(1)}$ and $ \Psi^{(2)} $ with the same profile but having disjoint supports respectively on the left and on the right of the origin. Both $\Psi^{(1)}$ and $\Psi^{(2)}$ have an average momentum heading toward the origin, with equal absolute value. \\
At time $n$ the state will be the sum of the independent evolution of the two states making up the initial superposition: $\Psi(n) = \Psi^{(1)}(n) +\Psi^{(2)}(n) = \{x_{j}^{(1)}(n) + x_{j}^{(2)}(n)\,,\,y_{j}^{(1)}(n) + y_{j}^{(2)}(n)\}_{j=1}^{N}$. Usually, there will be points of the lattice where both the $\Psi^{i}(n)$ have components different from zero. In those points the position probability will be obviously not the sum of the probability in the states $\Psi^{i}(n)$ being in general
\[  (x_{j}^{(1)}(n) + x_{j}^{(2)}(n))^{2} + (y_{j}^{(1)}(n) + y_{j}^{(2)}(n))^{2} \neq [(x_{j}^{(1)}(n))^{2} + (y_{j}^{(1)}(n))^{2}] + [(x_{j}^{(2)}(n)^{2}  + (y_{j}^{(2)})^{2}(n)] \]

\vspace{.3cm}
\noindent
In the following figure  we plotted the position probabilities at each point of the lattice for three successive times computed solving the recurrence equation (\ref{Sch})
\begin{figure}[htbp] \label{int}\centering 
\includegraphics[width=55mm]{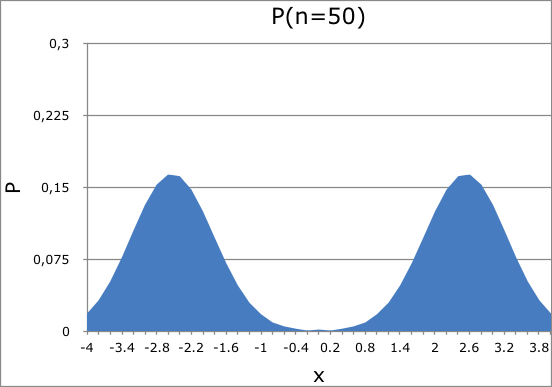}
\includegraphics[width=55mm]{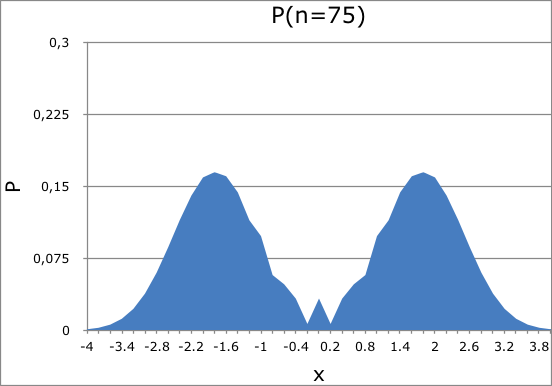}
\includegraphics[width=55mm]{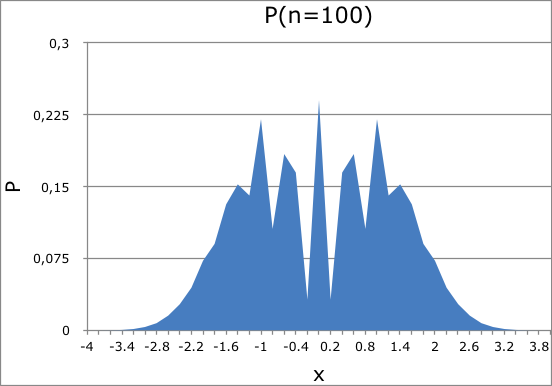}
        \caption{\emph{The evolution of position probability of a free quantum particle  when the initial distribution consists of two distinct ``bumps'' with opposite average momentum, respectively at the right and the left of the origin. The probability to be in origin is not given by the sum of the two independent evolutions.}} 
 \end{figure}
 
 \vspace{.3cm}
\noindent
The comparison between the evolution of the probability density of a quantum particle and the the one of a brownian particle, makes clear the deep difference existing between the to cases. In the former case it is impossible to characterize the initial superposition as ``either the particle is on the right and is moving to the left or it is on the left and is moving to the right ''. This fact prevents the possibility to define any sort of conditional probability in the quantum case and to find an evolution equation for the position probability density. 

\begin{figure}[htbp] \centering 
\includegraphics[width=70mm]{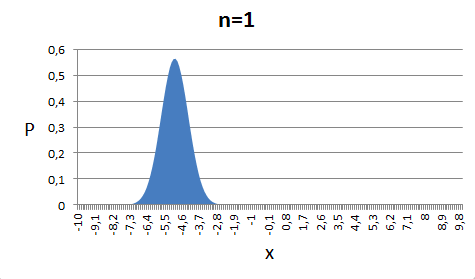}
\includegraphics[width=70mm]{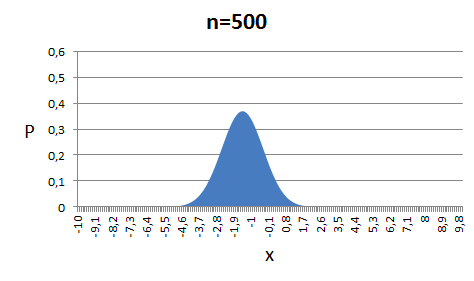}
\includegraphics[width=70mm]{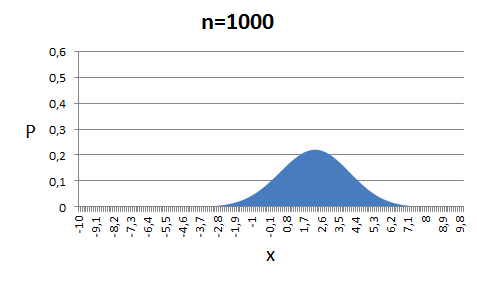}
        \caption{\emph{The evolution of the position probability of a free quantum particle shows the wave packet spreading. }} 
        \label{spread}
 \end{figure}

\vspace{.3cm}
\noindent
Others key aspects of quantum evolution can be presented using the numerical computation procedure described above. In the following we list some examples.\\
Investigating the evolution of a free wave packet it is possible to check that Schr\"odinger's equation is dispersive and to analyze the connection between the dispersive phenomenon and the indeterminacy in the initial momentum.  Indeed the {\figurename\ref{spread}} shows \emph{the spread in time of the wave packet }.
 
\begin{figure}[htbp] \centering 
\includegraphics[width=60mm]{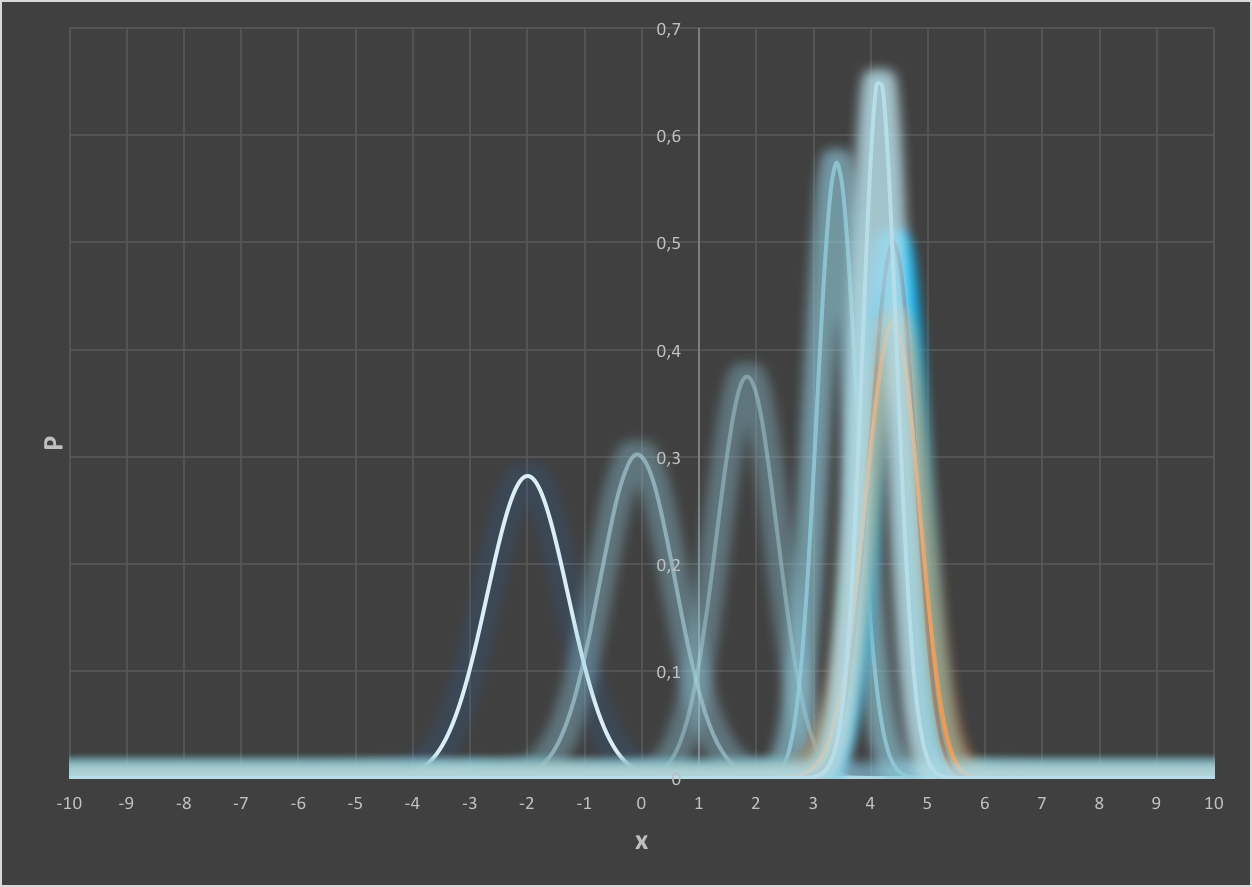}
\includegraphics[width=60mm]{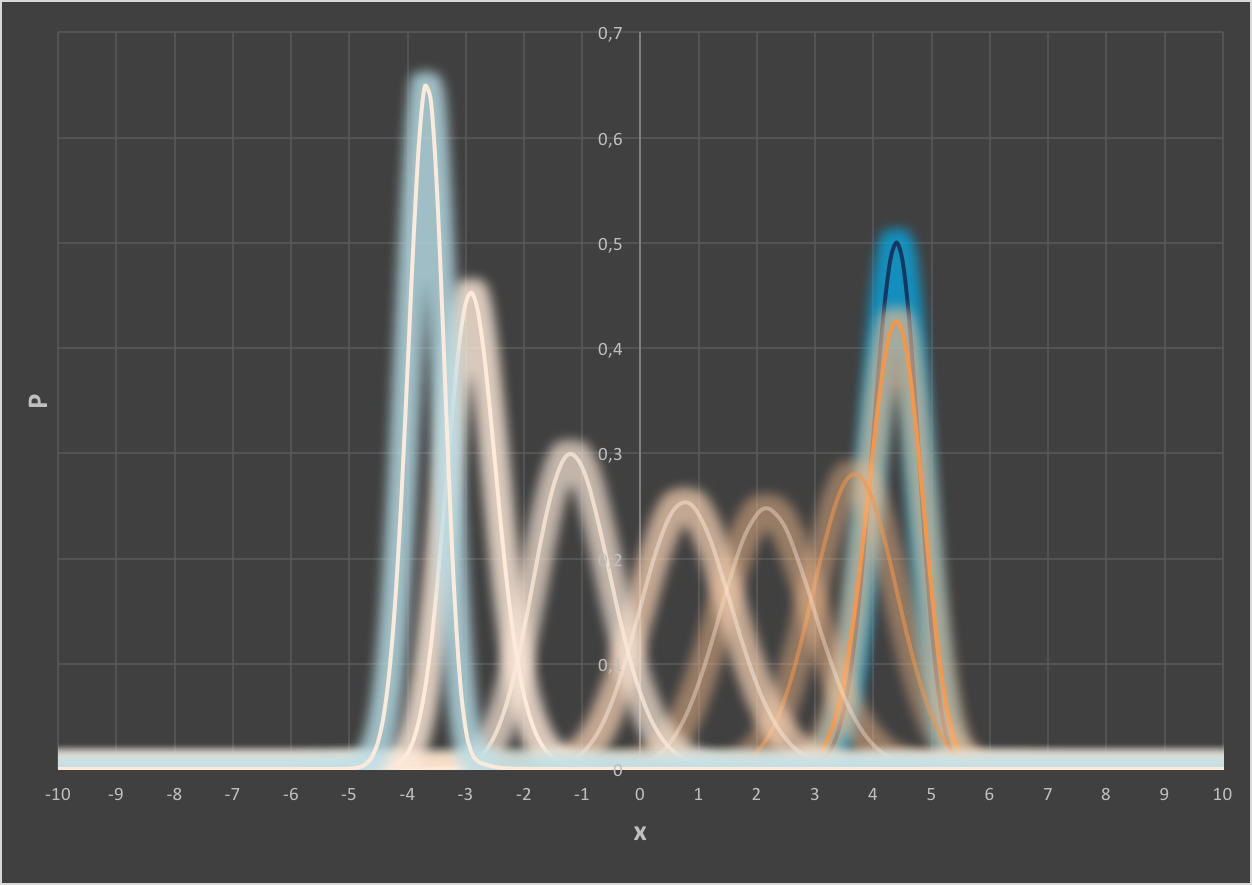}
\includegraphics[width=60mm]{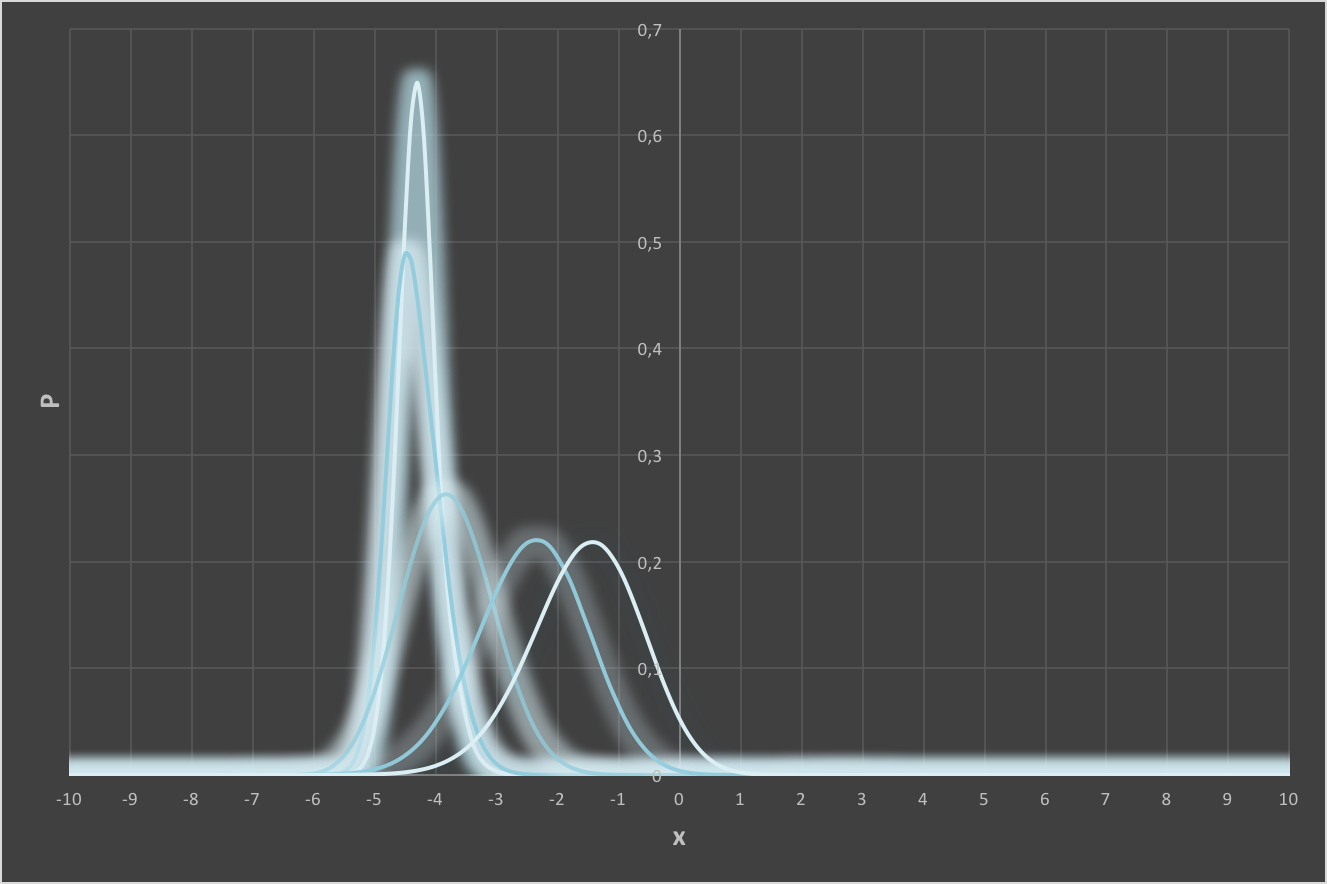}
        \caption{\emph{For a coherent state, the evolution shows the position probability confined in a harmonic trap.}} 
        \label{Co}
 \end{figure}

\noindent
It is also possible to examine the case of a \emph{quantum harmonic oscillator}.   Including the elastic potential $V_i=-Kx_i^2$ in equation \ref{Sch}, it is possible to verify, for different initial conditions, the existence of stationary solutions as well as the existence of travelling solutions   (``almost classic'' states).  
In figure {\figurename\ref{Co}} it is shown the evolution of a \emph{coherent state}. One can notice that  the probability evolves like the position of a classical particle oscillating in a harmonic trap.  

\section{Comments and Conclusions}
Our aim was to plan an unified strategy to present theoretical and computational aspects of classical and modern physics evolution equations to high school students. The enormous technical difficulties one encounters in dealing with complex system dynamics, using the whole equipment of an analytical theory, always prevented to insert these subjects in any high school syllabus. \\
Presenting phenomenology and interpretations of brownian motion in section 4 we mentioned that very powerful physics simulation softwares are nowadays available and widely used in scientific teaching. Students can select "objects" having specific geometric and dynamical properties,  evolution rules, interactions among objects and generic initial conditions. Programs then compute and display the successive evolution even for systems of great complexity. They are important resources allowing students to examine in their own way the predictions of a theoretical model.  Unfortunately the sophisticated computational protocols behind the object oriented programming remains concealed and never revealed to students not experienced in modern programming languages.\\
Our attempt has been to find a way to bridge the just mentioned gap using simplified forms of numerical calculus in order to make as clear as possible how computational methods work. We aimed to allow students to understand, evaluate the effectiveness and manage to find solutions to the evolution equations of physics.\\
The proposal outlined in the paper was presented in the university course " Didactics of Physics"  addressed to  second year students in physics and mathematics and in a training course for high school science teachers.  
We working so that more details along with all the simulations and animations made using the spreadsheet as a simplified calculation solver of recurrence equations are soon available on a website.

\end{document}